\documentclass[final,3p,times]{elsarticle}
\usepackage{amsfonts}
\usepackage{mathrsfs}
\usepackage{amsmath}
\usepackage{xcolor}
\usepackage{amssymb}
\usepackage{bm}
\usepackage{slashed}
\usepackage{graphicx}
\usepackage{subfigure}
\usepackage{verbatim} 
\usepackage[normalem]{ulem}
\usepackage[linktocpage=true,bookmarks=true,bookmarksnumbered=true,breaklinks=true,pdfpagemode=Fullscreen,pdfstartview=FitBH]{hyperref}

\numberwithin{equation}{section}
\allowdisplaybreaks[4]

\definecolor{gesfpurple}{rgb}{0.47,0.19,0.42}

\definecolor{gesflanse}{rgb}{0.00,0.50,0.50}

\definecolor{gesfblue}{rgb}{0.08,0.42,0.76}

\definecolor{gesfred}{rgb}{1,0,0}

\definecolor{gesfwhite}{rgb}{1,1,1}

\definecolor{gesfblack}{rgb}{0,0,0}

\newcommand{\geqn}[1]{\hypersetup{linkcolor=blue}(\ref{#1})\hypersetup{linkcolor=blue}}
\newcommand{\gfig}[1]{{\hypersetup{linkcolor=violet}Fig.\,\ref{#1}\hypersetup{linkcolor=blue}}}

\newcommand{\fr}[2]{\mbox{$\frac{\,{#1}\,}{#2}$}}

\def\bge{\begin{equation}}
\def\ede{\end{equation}}
\def\bga{\begin{aligned}}
\def\eda{\end{aligned}}
\def\bgp{\begin{pmatrix}}
\def\edp{\end{pmatrix}}
\def\bgs{\begin{subequations}}
\def\eds{\end{subequations}}
\newcommand{\beq}{\begin{equation}}
\newcommand{\eeq}{\end{equation}}
\newcommand{\bq}{\begin{equation}}
\newcommand{\eq}{\end{equation}}
\newcommand{\ba}{\begin{array}}
\newcommand{\ea}{\end{array}}
\newcommand{\beqa}{\begin{eqnarray}}
\newcommand{\eeqa}{\end{eqnarray}}
\newcommand{\beqs}{\begin{subequations}}
\newcommand{\eeqs}{\end{subequations}}

\def\[{\left[}
\def\]{\right]}
\def\({\left(}
\def\){\right)}

\def\leqq{\leqslant}

\def\la{\lambda}

\def\fr{\frac}

\def\End{\end{document}}

\begin{document}

\begin{frontmatter}

\setcounter{footnote}{0}
\renewcommand{\thefootnote}{\fnsymbol{footnote}}

\title{{\bf Flavor Structure of the Cosmic-Ray Electron/Positron Excesses at DAMPE}}

\author{
{\sc\large Shao-Feng Ge}\,\footnote{gesf02@gmail.com}$^{a,b}$,~~~
{\sc\large Hong-Jian He}\,\footnote{hjhe@tsinghua.edu.cn}$^{c,d,e,f}$,~~~
{\sc\large Yu-Chen Wang}\,\footnote{wang-yc15@mails.tsinghua.edu.cn}$^{e}$
\vspace*{3mm}
}

\address{
$^a$\,Kavli IPMU (WPI), UTIAS, The University of Tokyo, Kashiwa, Chiba 277-8583, Japan.
\\[1mm]
$^b$\,Department of Physics, University of California, Berkeley, CA 94720, USA.
\\[1mm]
$^c$\,T.~D.~Lee Institute, Shanghai 200240, China.
\\
$^d$\,School of Physics and Astronomy,
Shanghai Jiao Tong University, Shanghai 200240, China.
\\[1mm]
$^e$\,Institute of Modern Physics, Tsinghua University, Beijing 100084, China.
\\[1mm]
$^f$\,Center for High Energy Physics, Peking University, Beijing 100871, China.
\vspace*{-8mm}
}

\begin{abstract}
The Dark Matter Particle Explorer (DAMPE) satellite detector
announced its first result for measuring the
cosmic-ray electron/positron (CRE) energy spectrum up to 4.6\,TeV,
including a tentative peak-like event excess at $(1.3-1.5)$TeV.
In this work, we uncover a significant hidden excess
in the DAMPE CRE spectrum over the energy range $(0.6-1.1)$TeV,
which has a non-peak-like structure.
We propose a new mechanism to explain this excess by a set of
1.5\,TeV $\mu^\pm$ events with subsequent decays into $e^\pm$ plus neutrinos.
For explaining this new excess together with the peak excess around 1.4\,TeV,
we demonstrate that the {\it flavor structure} of the original lepton final-state produced
by dark matter (DM) annihilations (or other mechanism) should have a composition ratio
$N_e \!:\! (N_\mu \!+\!\fr{1}{6}N_\tau) = 1\!:\!y$\,,\,
with $\,y \simeq 2.6\!-\!10.8\,$.\,
For lepton portal DM models, this puts nontrivial constraint
on the lepton-DM-mediator couplings
$\,\lambda_e^{}\! : (\lambda_\mu^4 \!+\!\frac{1}{6}\lambda_\tau^4)^{\frac{1}{4}}
 =1:y^{\frac{1}{4}}\,$
with a narrow range $\,y^{\frac{1}{4}}\!\simeq\! 1.3\!-\!1.8\,$.
\\[1.5mm]
\\[1.5mm]
Phys.\ Lett.\ B\,781 (2018) 88 $[$arXiv:1712.02744 [astro-ph.HE]$]$.
\end{abstract}


\end{frontmatter}

\graphicspath{{figs/}}

\renewcommand{\thefootnote}{\arabic{footnote}}
\setcounter{footnote}{0}

\section{Introduction}
\label{sec:1}

Measuring high energy cosmic ray electrons and positrons (CRE)
is important for probing the nearby galactic sources\,\cite{source}
and the possible observation of dark matter (DM) annihilations\,\cite{inDM}.
The cosmic ray electron/positron spectrum has been probed
up to TeV energy scales by the ground-based and space-borne
experiments such as HESS\,\cite{HESS}, VERITAS\,\cite{VERITAS},
FermiLAT\,\cite{Fermi}, AMS-02\,\cite{AMS}, and CALET\,\cite{CALET}.
These provide important means for the indirect DM detection.

\vspace*{1mm}

The Dark Matter Particle Explorer (DAMPE) satellite detector\,\cite{DAMPE}
was launched at the end of 2015 and is optimized for measuring cosmic $e^\pm$ rays
and $\gamma$-rays up to about 10\,TeV energy. After 530\,days of data-taking,
the DAMPE collaboration newly announced\,\cite{DAMPE2017}
its first result of detecting the
cosmic-ray electron/positron energy spectrum from 25\,GeV up to 4.6\,TeV.
The main part of this spectrum can be fitted by a smoothly broken power-law
model and shows a spectral break around 0.9\,TeV, which confirms the similar
evidence found by HESS\,\cite{HESS}. The DAMPE CRE spectrum also indicates
a tentative peak-like event excess around 1.4\,TeV.
This excess has triggered wide interests and attempts
to the possible physics implications\,\cite{DAMPE-imply}-\cite{models},
either from the conventional astrophysical sources
(such as pulsars and supernova remnants)
or from dark matter annihilation into $e^+e^-$ events,
as well as building various lepton-related DM models.

\vspace*{1mm}

In this work, we uncover a significant new excess over the energy region
$(0.6-1.1)$TeV on the left-hand-side of the peak bin $(1.3-1.5)$TeV.
We propose to explain this intriguing new excess
with 1.5\,TeV $\mu^\pm$ events which are produced
together with the 1.5\,TeV $e^\pm$ peak events and subsequently decay
into $e^\pm$ plus neutrinos.
For a possible 1.5\,TeV $\tau$-component in the original CRE source,
its leading decay contribution to the DAMPE detection is much smaller than muon,
but has a similar energy distribution.
To explain this new excess together with the peak,
we find that the {\it flavor structure} of the original lepton final-state
produced by the DM annihilations (or other mechanism)
from DM annihilations (or other mechanism) receives a nontrivial constraint.
Although confirming this new excess and/or the tentative 1.4\,TeV peak structure
would require more data-taking from DAMPE and other CRE experiments,
these encouraging clues are important and deserve further investigations.

\vspace*{1mm}

This paper is organized as follows. In Sec.\,\ref{sec:2},
we first inspect the
DAMPE CRE spectrum and uncover a new event excess over
the energy region $(0.6-1.1)$TeV.
Then, we fit the CRE background data points
by treating this new excess and the 1.4\,TeV peak excess as signals.
Our new fit gives a fairly smooth background curve and has much lower $\chi^2$
than the naive fit (treating all data points as backgrounds and
with much poorer quality).
We further examine the improved sensitivity
of the continued DAMPE running up to $5\!-\!6$ years,
which will help to pin down the new excess and peak excess.
In Sec.\,\ref{sec:3}, we will study the decays of 1.5\,TeV $\mu^\pm$ events
into $e^\pm$ plus neutrinos,
which are produced at the same time as the 1.5\,TeV $e^\pm$ peak events.
We show that the decay contributions of the 1.5\,TeV $\mu^\pm$ events
to the DAMPE CRE spectrum can explain very well the new excess over
the region $(0.6-1.1)$TeV.
In Sec.\,\ref{sec:4}, we perform a fit to include
both the decay-contribution of 1.5\,TeV $\mu^\pm$ events
and the 1.5\,TeV $e^\pm$ events produced together in a nearby clump
or subhalo source. We derive the original flavor composition of the
CRE spectrum detected by DAMPE.
We further analyze the physics implications of the
original flavor structure of the DAMPE CRE spectrum.
Finally, we conclude in Sec.\,\ref{sec:5}.

\section{Backgrounds and New Excess Off the Peak}
\label{sec:2}

\begin{figure}[t]
\centering
\includegraphics[height=7cm,width=8.7cm]{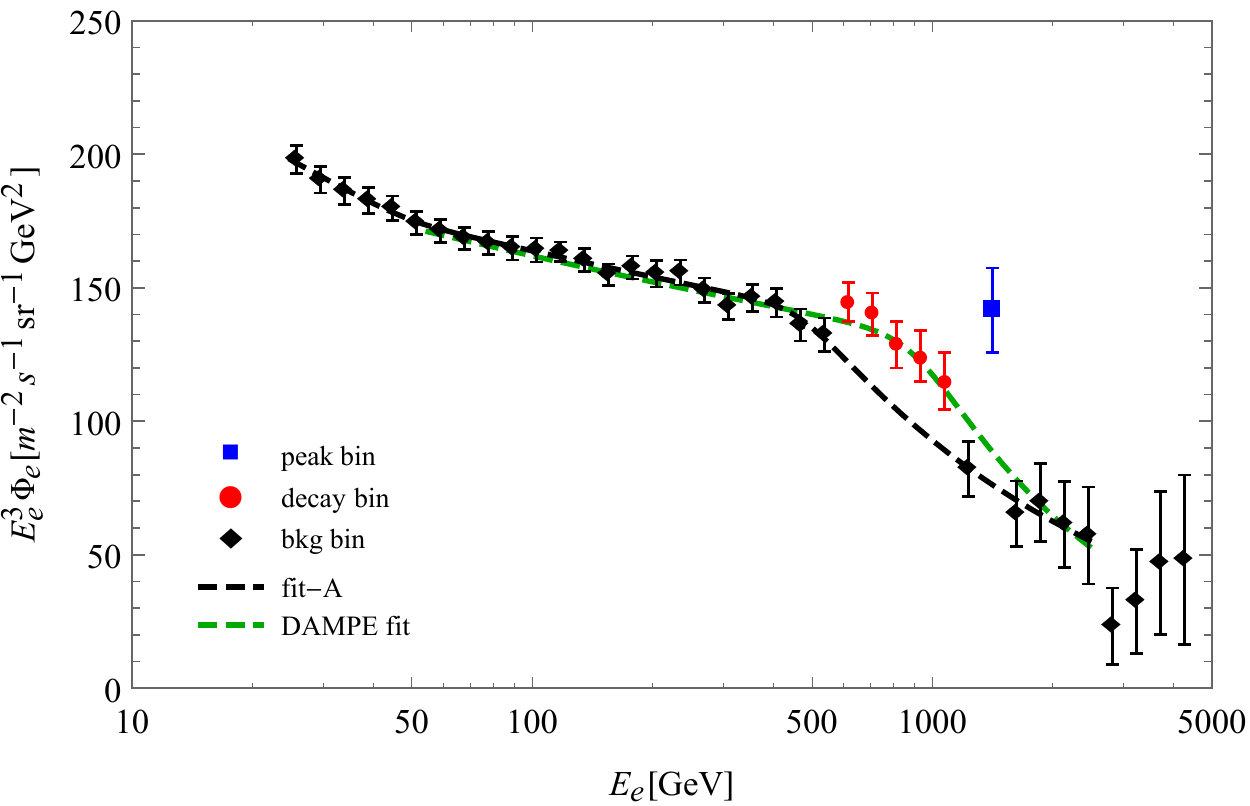}
\vspace*{-2mm}
\caption{Fitting the background CRE spectrum (black bins) at DAMPE,
as shown by the black dashed curve for $25\,\text{GeV}\!-\!2.6\,$TeV region.
The blue peak bin at $(1.3\!-\!1.5)$TeV is more than $3\sigma$ above the background curve
and each of the 5 red bins in the region $(0.6-1.1)$TeV deviates from the background curve
by $\gtrsim\!2\sigma$, so they are not included in the background fit.
The green dashed curve reproduces the original DAMPE fit\,\cite{DAMPE2017}
over $55\text{GeV}\!-2.6$TeV (including both the 5 red bins and the blue peak bin),
which has much poorer fitting quality as discussed in the text.
}
\label{fig:fit-BKG}
\label{fig:1}
\end{figure}

We replot all the DAMPE data points with $\pm 1\sigma$ errors\,\cite{DAMPE2017} in
Fig.\,\ref{fig:1}. This is much clearer than the original plot (Fig.2)
of DAMPE\,\cite{DAMPE2017} which displayed three other experiments
(HESS, Fermi-LAT, AMS-02)
altogether for comparison, where the HESS and Fermi-LAT results
have significant overlap with the DAMPE data points around $(0.35-2)$TeV
and make the precise feature of DAMPE points less clear in this energy region.

\vspace*{1mm}

The DAMPE data points presented in the current Fig.\,\ref{fig:1}
appear distinctive. From this, we observe that the DAMPE data points
exhibit another rather intriguing structure
on the left-hand-side of the peak $(1.3-1.5)$TeV.
The energy region of $(0.616-1.07)$TeV
contains five consecutive data points (marked in red color), which
all lie above the expected positions (based on fitting the rest of
the background points).
These five red data points are distinctive
and form a non-peak-like {\it new excess}
in addition to the peak bin (marked in blue).
In particular, we compare these five red points with the rest
and find {\it clear jumps} above their neighboring points (marked in black color)
by $\gtrsim\!2\sigma$ deviations.

\vspace*{1mm}

To make this feature fully clear,
we first fit the background bins
(marked by black color in the current Fig.\,\ref{fig:1}),
without including the five red data points over $(0.616\!-\!1.07)$TeV
and the blue peak point at $(1.3\!-\!1.5)$TeV.
For this background fit,
we use the double-broken power-law form\,\cite{PL},
\begin{equation}
\label{eq:DBPL}
  \Phi_{\text{bkg}}^{} \,=\,
\Phi_0^{}\(\!\frac{\,\text{GeV}\,}{\,E_e^{}\,}\!\)^{\!\gamma}_{}
\left[
1\!+\! \left(\! \frac{\,E_{\rm br 1}}{E_e^{}} \!\right)^{\!\delta}
\right]^{\frac {\,\Delta\gamma_1} \delta}\!
\left[1
\!+\! \left(\! \frac{E_e^{}}{\,E_{\rm br2}\,} \!\right)^{\!\delta}
\right]^{\!\frac{\,\Delta\gamma_2} \delta} \!\!,~~~~
\end{equation}
where the flux $\,\Phi\,$ is defined as the number of events per unit area,
per unit solid angle, per unit time, and per unit energy width.
According to the recent results of Fermi-LAT \cite{Fermi} and DAMPE \cite{DAMPE2017},
we can set the first break
$\,E_{\rm br 1}^{} \!\!= 50$\,GeV\, and the
sharpness parameter $\,\delta = 10$\,.\,

\vspace*{1mm}

Then, we perform a minimal $\chi^2$ fit
\beqa
\label{eq:chi2}
\chi^2 \,=\,
  \sum_j
\left[
  \frac{\,\Phi_e(E_j^{}) - \mathcal O_j^{}\,}
       {\Delta \mathcal O_j^{}}
\right]^2 \,,
\eeqa
for all the background bins (black color).
In Eq.\eqref{eq:chi2}, each experimental central value
and its error are expressed as
$\mathcal O_j^{}$ and $\Delta \mathcal O_j^{}$, respectively.
We make two background fits A and A$'$, where we exclude
the blue peak bin at $(1.3\!-\!1.5)$ TeV and the 5 red bins over $(0.6-1.1)$TeV,
and will treat them as signal bins.
The fit-A covers the energy range $25\text{GeV}\!-\!2.6\text{TeV}$ and
the fit-A$'$ spans over $25\text{GeV}\!-\!4.6\text{TeV}$.
Since the last 4 bins in $(2.6\!-\!4.6)$TeV region have rather low event counts
($\leqq\!4$) and thus much larger statistical errors, they were also excluded
from the fit of DAMPE Group\,\cite{DAMPE2017}. So fit-A$'$ is not our focus
and we mention it only as a reference.

\vspace*{1mm}

Our fit-A gives:
$\,\Phi_0^{} = 246^{+15}_{-14}\text{GeV}^{-1}\text{m}^{-2}\text{s}^{-1}\text{sr}^{-1}$,\,
$\gamma =3.09\pm 0.01$,
$(\Delta \gamma_1^{},\,\Delta \gamma_2^{}) =
(0.095^{+0.045}_{-0.045},\, -0.48^{+0.08}_{-0.11})$,\,
and $\,E_{\rm br 2}^{} = 471^{+93}_{-59}\,\mbox{GeV}$.
We obtain the minimal $\,\chi^2=3.95$\, and the degrees of freedom
(d.o.f.) is 23, which give $\,\chi^2\!/\text{d.o.f.}=3.95/23=0.172$.
Then, we use the best fit of fit-A and present the fitted
background CRE spectrum by the black dashed curve in Fig.\,\ref{fig:1}.
As a reference, our fit-A$'$ gives:
$\Phi_0^{} = (247\pm 14)\,
\text{GeV}^{-1}\text{m}^{-2}\text{s}^{-1}\text{sr}^{-1}$,\,
$\gamma =3.09\pm 0.01$,
$(\Delta \gamma_1^{},\,\Delta \gamma_2^{}) =
(0.093^{+0.045}_{-0.045},\,-0.56^{+0.09}_{-0.11})$,\,
and $\,E_{\rm br 2}^{}\! = (493^{+126}_{-59})\,$GeV.
This fit-A$'$ has $\,\chi^2=7.60$\,,  $\,\text{d.o.f.}=27$\,,\, and thus
$\,\chi^2/\text{d.o.f.}=7.60/27=0.281$\,.\,
We see that the fitting quality of fit-A is better than fit-A$'$, as expected.

\vspace*{1mm}

For comparison, we perform another background fit-B
for the black bins over the energy range $55\,\text{GeV}\!-\!2.6\,$TeV,
by using the single broken power-law formula\,\cite{DAMPE2017},
\beqa
\label{eq:SBPL}
\Phi_{\text{bkg}}^{}
\,=\,\Phi_0^{}\(\!\frac{\,100\text{GeV}\,}{E_e^{}}\!\)^{\gamma_1^{}}\!
\left[1\!+\(\!\frac{E_e^{}}{\,E_{\text{br}}^{}\,}\!\)^{\!(\gamma_2^{}\!-\gamma_1^{})
       /\Delta}\right]^{-\Delta}\!,~~~~~~
\eeqa
with the choice of the smoothness parameter $\Delta =0.1$\,.\,
This formula was used by DAMPE Group\,\cite{DAMPE2017} for fitting the
same energy region $55\text{GeV}\!-2.6$TeV.
Our fit-B gives:
$\Phi_0^{}=(1.64\pm 0.014)\!\times\!10^{-4}\,
 \text{GeV}^{-1}\text{m}^{-2}\text{s}^{-1}\text{sr}^{-1}$,\,
$(\gamma_1^{},\,\gamma_2^{}) = (3.08^{+0.02}_{-0.02},\, 3.62^{+0.12}_{-0.09})$,\,
and $\,E_{\text{br}}^{}=518^{+112}_{-77}\,$ GeV.
This fit-B has
$\,\chi^2/\text{d.o.f.}$ $=3.48/18=0.193$\,,\,
which is quite similar to our fit-A.

\vspace*{1mm}

Our background fits (A,\,B) have excluded the 5 red bins
and the blue peak bin in Fig.\,\ref{fig:1},
while the original DAMPE fit\,\cite{DAMPE2017} included them.
For comparison,
we have reproduced the DAMPE fit (without nuisance parameters)
for all data points in the energy range
$55\text{GeV}\!-\!2.6$TeV,
and obtain $\,\chi^2/\text{d.o.f.}=22.5/24=0.939$\,,\,
where we have combined the statistical and systematical errors.
Our total $\,\chi^2=22.5\,$ is fairly close to the DAMPE fit
$\chi^2=23.3$ \cite{DAMPE2017}
(with $\text{d.o.f.}=18$ including nuisance parameters)\,\cite{foot1}.
We show the best fit as the green dashed curve in Fig.\,\ref{fig:1},
which agrees very well with the original DAMPE fit (cf.\
the red dashed curve in Fig.\,2 of Ref.\,\cite{DAMPE2017}).
This green curve appears un-natural because both the bottom bins on the two sides
of the peak bin $(1.3-1.5)$TeV lie significantly below the green dashed curve,
but they match perfectly well with our new background fits (black curves).
Then, we redo the DAMPE fit by taking out the peak bin and
obtain $\,\chi^2/\text{d.o.f.}=10.1/23=0.440$,\,
which becomes smaller by more than a factor 2 than before.
We see that the total $\chi^2=10.1$ decreases from the reproduced
DAMPE fit of $\chi^2=22.5$ by a large amount $\,\Delta\chi^2=12.4\,$,\,
while the fitting degrees of freedom (d.o.f.) is changed by just one.
Next, we further redo the DAMPE fit by taking out both the peak bin
and the 5 red bins. This is the same as our fit-B
where we treat both the 5 red bins and the peak bin as signals.
As shown earlier, our fit-B gives
$\,\chi^2\!/\text{d.o.f.}=3.48/18=0.193$,\,
which is smaller than the reproduced DAMPE fit of
$\,\chi^2/\text{d.o.f.}$\,($=22.5/24=0.939$)\, by a factor 4.9
(with $\,\Delta\chi^2\simeq 19.0\,$), and
is smaller than the above fit of excluding the peak bin only and treating
the 5 red bins as backgrounds ($\chi^2\!/\text{d.o.f.}=10.1/23=0.440$)
by a factor of 2.3 (with $\Delta\chi^2\simeq 6.63\,$).
The reproduced DAMPE fit of
naively treating both the peak bin and the 5 red bins as
backgrounds has $\,\chi^2\!/\text{d.o.f.}=22.5/24=0.939$,\,
which is also larger than our fit-A
($\chi^2\!/\text{d.o.f.}=3.95/23 = 0.172$)
by a factor of 5.5, and has
$\,\Delta\chi^2\simeq 18.6\,$.\,

\vspace*{1mm}

The DAMPE detector is expected to run up to $5\!-\!6$ years or even longer
and will accumulate about four times of the current data
set\,\cite{DAMPE}\cite{DAMPE2017}, which will reduce
the statistical errors by half.
Hence, assuming the same central values as the present,
we redo fit-A for future DAMPE running up to 6\,years, which will be
called fit-C hereafter. We find that the quality of our fit-C will become
$\chi^2\!/\text{d.o.f.}=$ $5.93/23=0.258$, while the naive fit of treating
all bins over $25\,\text{GeV}\!-\!2.6\,$TeV as backgrounds will have
$\chi^2\!/\text{d.o.f.}=62.4/29=2.15$,\, which is higher than our fit-C
by a large factor 8.3\,.\,
Here this naive fit has a large total $\,\chi^2 = 62.4\,$,\,
and is higher than our fit-C by a big amount
$\,\Delta\chi^2 = 56.5\,$.\,
Hence, the future DAMPE runs up to 6\,years will much help to pin down
the new excess over $(0.6\!-\!1.1)$TeV
in addition to the 1.4\,TeV peak structure.

\vspace*{1mm}

\gfig{fig:1} presents the fit-A of the background bins (black color)
as the black dashed curve, where we use the
double-broken power-law formula \eqref{eq:DBPL} and set all parameters
to their best fit values.
This new background curve is fairly smooth and
has better quality than the naive fit of treating all bins
as backgrounds (cf.\ the green dashed curve in \gfig{fig:1}),
as explained above. 
It also makes the two bottom bins on both sides of the peak bin appear
{\it fully natural and match very well with our background curve.}
\gfig{fig:1} shows that above the background curve of fit-A,
the DAMPE data exhibit two distinctive excess structures:
(i)~one is the previously noticed peak excess
at $(1.3\!-\!1.5)$TeV ($\gtrsim\!3\sigma$)
marked by blue color; and
(ii)~another is the non-peak-like new excess in the
$(0.6\!-\!1.1)$\,TeV region (marked in red color and
$\gtrsim\!2\sigma$ in every red bin).

\begin{table*}[t]
\centering
\begin{tabular}{c|c|c|c|c|c}
\hline\hline
& \multicolumn{3}{c|}{} & \multicolumn{2}{c}{}
\\[-3mm]
\hspace*{-2mm} Energy Region & \multicolumn{3}{c|}{$\chi^2$\,/\,d.o.f.}
& \multicolumn{2}{c}{Signal Significance}
\hspace*{-1.5mm}
\\
\cline{2-6}
\hspace*{-2mm} (TeV) & Naive\,BKG\,Fit & \!\!BKG\,Fit\,(no\,peak\,bin)\!\!
      & Our\,BKG\,Fit & Peak\,Excess & New\,Excess
\hspace*{-1.5mm}
\\
\hline
& & & & &
\\[-3mm]
\hspace*{-2mm}
0.025$-$2.6 & \!\!$22.8/29\!=\!0.785$\!\! & $10.3/28\!=\!0.369$
& \!$3.95/23\!=\!0.172$\,(fit-A)\!
& 4.1$\sigma$\,(4.0$\sigma$) & 6.6$\sigma$\,(2.9$\sigma$)
\hspace*{-1.5mm}
\\
& & & & &
\\[-3.2mm]
\hspace*{-2mm}
0.055$-$2.6 & \!\!$22.5/24\!=\!0.939$\!\! & $10.1/23\!=\!0.440$
& \!$3.48/18\!=\!0.193$\,(fit-B)\!
& 4.1$\sigma$\,(3.9$\sigma$) & 6.0$\sigma$\,(3.0$\sigma$)
\hspace*{-1.5mm}
\\
& & & & &
\\[-3.2mm]
\hspace*{-2mm}
0.025$-$2.6\,(for\,6y)\!\! & $62.4/29\!=\!2.15$~\,
& $23.2/28\!=\!0.829$
& \!$5.93/23\!=\!0.258$\,(fit-C)\!
& 7.3$\sigma$\,(7.2$\sigma$) & 10.8$\sigma$\,(8.2$\sigma$)
\hspace*{-1.5mm}
\\[1mm]
\hline\hline
\end{tabular}
\caption{%
Comparison of different background (BKG) fits.
For the energy range (0.025$-$2.6)TeV,
we use the double-broken power-law formula \eqref{eq:DBPL} for fit, while for
the range (0.055$-$2.6)TeV, we apply the single-broken power-law formula
\eqref{eq:SBPL} for fit.
In the 2nd column, ``Naive BKG Fit" includes all data bins
into the BKG fit. In the 3rd column, ``BKG Fit\,(no\,peak\,bin)"
takes out the blue peak bin from the BKG fit.
In the 4th column, ``Our BKG Fit" takes out both the blue peak bin and
the 5 red bins from our current BKG fit.
The last two columns show the local signal significance of ``Peak Excess"
(for the blue peak bin) and ``New Excess" (for the 5 red bins) based on the
corresponding BKG fits (A,\,B,\,C) in the 4th column; here the significance number
in each $(\cdots)$ correspond to statistically fluctuating the break parameter
$E_{\text{br2}}^{}$ or $E_{\text{br}}^{}$ to its 90\% upper limit.
In the last row, we make fit-C for future DAMPE running up to
6\,years (6y). 
This comparison well motivates treating the peak bin and 5 red bins as new signals
beyond the BKG fits. Even in the Naive BKG Fit, we find that the peak bin has a
local significance of $(3.5\sigma,\,3.3\sigma,\,3.3\sigma)$ for the energy
regions (0.025$-$4.6,\,0.025$-$2.6,\,0.055$-$2.6)TeV, which agrees to the
literature\,\cite{xfit}; while with this Naive BKG Fit we find that
the DAMPE 6-year running increases the peak significance to
$(6.2\sigma,\,6.0\sigma,\,5.9\sigma)$, which will well justify (or exclude) the
peak excess as new signal.
}
\label{tab:1}
\vspace*{-2mm}
\end{table*}

\vspace*{1mm}

According to the fit-A background curve in \gfig{fig:1},
we find that the 5 red bins contain a total background event number
$\,N_B^{}=1242.3\,$ and a total signal event number
$\,N_S^{}=285.7$,\, while the blue peak bin contains 50.0 background events
and 43.0 signal events. Hence, the 5 red bins have a much larger signal event
number than the peak bin. With these,
we can estimate the local signal significance of the 5 red bins and the peak bin
with the combined experimental errors,
where the statistical error always dominates over the systematical error in
each bin\,\cite{DAMPE2017}.
In each single bin, we may compute the signal significance
$\,\mathcal{Z}_j^{}=N_{Sj}^{}/\Delta N_j^{}\,$, where
$\,\Delta N_j^{}\,$ is the combined error for the $j$th-bin.
For the 5 red bins, we may estimate their total significance
$\,\mathcal{Z}_{\text{5rb}}^{}=(\sum_j\! \mathcal{Z}_j^2)^{1\!/2}$.\,
Using the fit-A background curve,
we find that the peak bin has a local signal significance
of $\,4.1\sigma$,\, and the 5 red bins has a signal significance of
$\,6.6\sigma$.\, 
In our fit-A, the second break parameter has the best fit
$\,E_{\text{br2}}^{}\!=471$GeV, and the upper limit
$\,E_{\text{br2}}^{}\!=738$\,GeV at 90\%\,C.L.
If we set $\,E_{\text{br2}}^{}\!=738\,$GeV,\,
we obtain $\,\chi^2\!/\text{d.o.f.}=6.66/23=0.289$.\,
So we see that the fit quality becomes worse than the best fit at
$\,E_{\text{br2}}^{}=$ $471$\,GeV
(with $\,\chi^2\!/\text{d.o.f.}=3.95/23=0.146$).
In this case, the 5 red bins contain the background and signal events
$\,(N_B^{},\,N_S^{})=(1409.9,\,118.1)$,\,
while the blue peak bin has 51.7 background events and 41.3 signal events.
So the 5 red bins have a signal significance of $2.9\sigma$,
while the peak bin has a signal significance of $4.0\sigma$.
This analysis shows that statistically fluctuating the break parameter
$E_{\text{br2}}^{}$ from the best-fit value to its 90\% upper limit
has little effect on the signal/background estimate in the peak bin,
but it can significantly affect the signal versus background division
in the 5 red bins. This is expected because the peak excess falls into
a single narrow bin which is insensitive to the background shape,
while the 5 red bins behave as a non-peak-like excess and are quite sensitive
to the shape of the background curve.
Nevertheless, it is impressive to note that even for a large value
$\,E_{\text{br2}}^{}\!=\!738$\,GeV (90\%\,C.L.), the signal significance still reaches
$3\sigma$ level for the 5 red bins.
We further verified that our fit-B also exhibits the similar features.
Namely, for the best fit values of fit-B, the peak excess has a significance
of $4.1\sigma$ and the new excess (5 red bins) has a significance of
$6.0\sigma$;\, when we statistically fluctuate $\,E_{\text{br}}^{}\,$
to its 90\% upper limit, the significance of peak excess slightly reduces to
$3.9\sigma$, while the new excess still has a $3.0\sigma$ significance.

\vspace*{1mm}

For further comparison, we summarize in Table\,\ref{tab:1} the results of
our background fits (A,\,B,\,C) as well as the naive background fit.
It shows that excluding the peak bin $(1.3\!-\!1.5)$TeV from the
background fits (3rd column)
always reduces the $\chi^2$ and $\chi^2$\!/d.o.f.\ by a factor $\gtrsim\!2$,
as compared to the naive background fits (2nd column).
By treating both the 5 red bins and peak bin as signal excesses,
the quality of our background (BKG) fits (4th column)
becomes better by another factor of $\sim\!\!2$\,.\,
In the 5th and 6th columns, we show the local signal significance of
the ``Peak Excess" (for blue peak bin) and ``New Excess" (for 5 red bins)
based on the best fits of the corresponding background fits
(A,\,B,\,C) in the 4th column;
while the value of significance in each $(\cdots)$ corresponds to statistically
fluctuating the break parameter $E_{\text{br2}}^{}$ or $E_{\text{br}}^{}$
to its 90\% upper limit.
In the last row, we present the fit-C for future DAMPE running up to
6\,years (6y) by assuming the same central values as the current data.

\vspace*{1mm}

Table\,\ref{tab:1} shows
that for our fits (A,\,B,\,C), the peak excess has a local significance
$(4.1\sigma,\, 4.1\sigma,\, 7.3\sigma)$, and setting the break parameter
$E_{\text{br2}}^{}$ or $E_{\text{br}}^{}$ be its 90\% upper limit only
slightly reduces the significance value as
$(4.0\sigma,\, 3.9\sigma,\, 7.2\sigma)$.\,
On the other hand, we can estimate that the new excess (the 5 red bins) has
a significance $(6.6\sigma,\, 6.0\sigma,\, 10.8\sigma)$ for fits (A,\,B,\,C),
while setting the break parameter be its 90\% upper limit significantly
reduces the significance value to
$(2.9\sigma,\, 3.0\sigma,\, 8.2\sigma)$.\,
Hence, although the new excess is non-peak-like and sensitive to the
shape parameter of the background curve, it is impressive that this new excess
still reaches $3\sigma$ level with the current DAMPE data\,\cite{DAMPE2017}
even if $E_{\text{br2}}^{}$ or $E_{\text{br}}^{}$
statistically fluctuates to its 90\% upper limit.
Furthermore, assuming the current central values of the peak bins and 5 red bins
remain, we estimate that the continued DAMPE running up to 6\,years will raise
the significance of the (peak, non-peak) excesses to
$(7.3\sigma,\, 10.8\sigma)$; while when the break parameter fluctuates to
its 90\% upper limit, the expected future significance of (peak, non-peak)
excesses will become $(7.2\sigma,\, 8.2\sigma)$,\, which are still well beyond
the $5\sigma$ level.

\vspace*{1mm}

In addition, even for the {naive background fit} including all bins
(2nd column of Table\,\ref{tab:1}),
we find the peak bin has a local significance of
$(3.5\sigma,\,3.3\sigma,\,3.3\sigma)$ for fitting the
energy regions (0.025$-$4.6, 0.025$-$2.6, 0.055$-$2.6)\,TeV,
which agrees to the literature\,\cite{xfit} and
is slightly lower than the peak significance $(4.4\sigma,\,4.1\sigma,\,4.1\sigma)$
given by our fits (A$'$,\,A,\,B).
For the future DAMPE running up to 6\,years,
we find that the naive background fit gives the increased peak significance
$(6.2\sigma,\,6.0\sigma,\,5.9\sigma)$ for the three energy ranges above,
which are well beyond the $5\sigma$ level
and will justify (or exclude) the peak excess as new signal.

\vspace*{1mm}

In summary, the above analysis and comparison convincingly show
that treating both the 5 red bins and
the peak bin as signals will significantly improve the quality
of the background fits,
hence this analysis is well motivated and instructive.
Our analysis (fit-C) further demonstrates that the improved sensitivity
by the future DAMPE runs will much help to pin down
both the new excess and the 1.4\,TeV peak structure.

\vspace*{1mm}

In passing, we note that this non-peak-like new excess over $(0.6\!-\!1.1)$TeV
is consistent with the recent data of another satellite experiment
Fermi-LAT\,\cite{Fermi}
(cf.\ the Fermi-LAT data points shown in Fig.\,2 of Ref.\,\cite{DAMPE2017}
for comparison).
In the high energy region ($E>1\,$TeV),  the energy resolution of Fermi-LAT is
about 20\% and is much larger than that of DAMPE [which is $(1\!-\!2)\%$].
The Fermi-LAT data have larger errors for $\,E>1$\,TeV,\, and are still consistent
with DAMPE data in this energy region, although the Fermi-LAT data points do not yet
exhibit a peak structure around $(1.3\!-\!1.5)$TeV due to much lower energy resolution.
The HESS data\,\cite{HESS} have similar behavior as DAMPE
(cf.\ Fig.\,2 of Ref.\,\cite{DAMPE2017}), but with much larger
systematical errors.

\vspace*{1mm}

Before concluding this section, we stress that
our above new observation and fits
(Fig.\,\ref{fig:1} and Table\,\ref{tab:1}) are very
suggestive and encouraging.
It intrigues us to conjecture that this non-peak-like new excess
over $(0.6\!-\!1.1)$TeV may be interconnected with
the peak structure around 1.4\,TeV.\, 
In the next section, we shall propose a new origin to explain this intriguing
non-peak-like structure.

\section{Decays of Muon Composition for New Excess}
\label{sec:3}

Inspecting the non-peak-like new excess over the energy range $(0.6-1.1)$TeV
of \gfig{fig:1}, we conjecture that it originates from the decays
of 1.5\,TeV muons.
These muon events were produced at the same time when the 1.5\,TeV $e^\pm$ events
were generated, say, during the DM annihilations in a nearby clump or subhalo.
Once $\mu^\pm$ events are produced, they will decay predominantly into $e^\pm$
via the 3-body channel with almost 100\% branching fraction\,\cite{PDG},
$\,\mu\to e \bar{\nu}^{}_e\nu_\mu^{}\,$.\,
A flying muon with 1.5\,TeV energy has a lifetime about \,0.031s\, and could
only travel about $9.3\!\times\!10^{6}$m.\, This distance is negligible
when compared to the nearby potential DM sources (typically within
$\sim\!1$kpc distance from the earth\,\cite{DAMPE-imply}).
All the $\mu^\pm$ events would convert into $e^\pm$ events, long before
they possibly reach the DAMPE detector.

\vspace*{1mm}

\begin{figure*}[t]
\centering
\includegraphics[height=6.5cm,width=8.1cm]{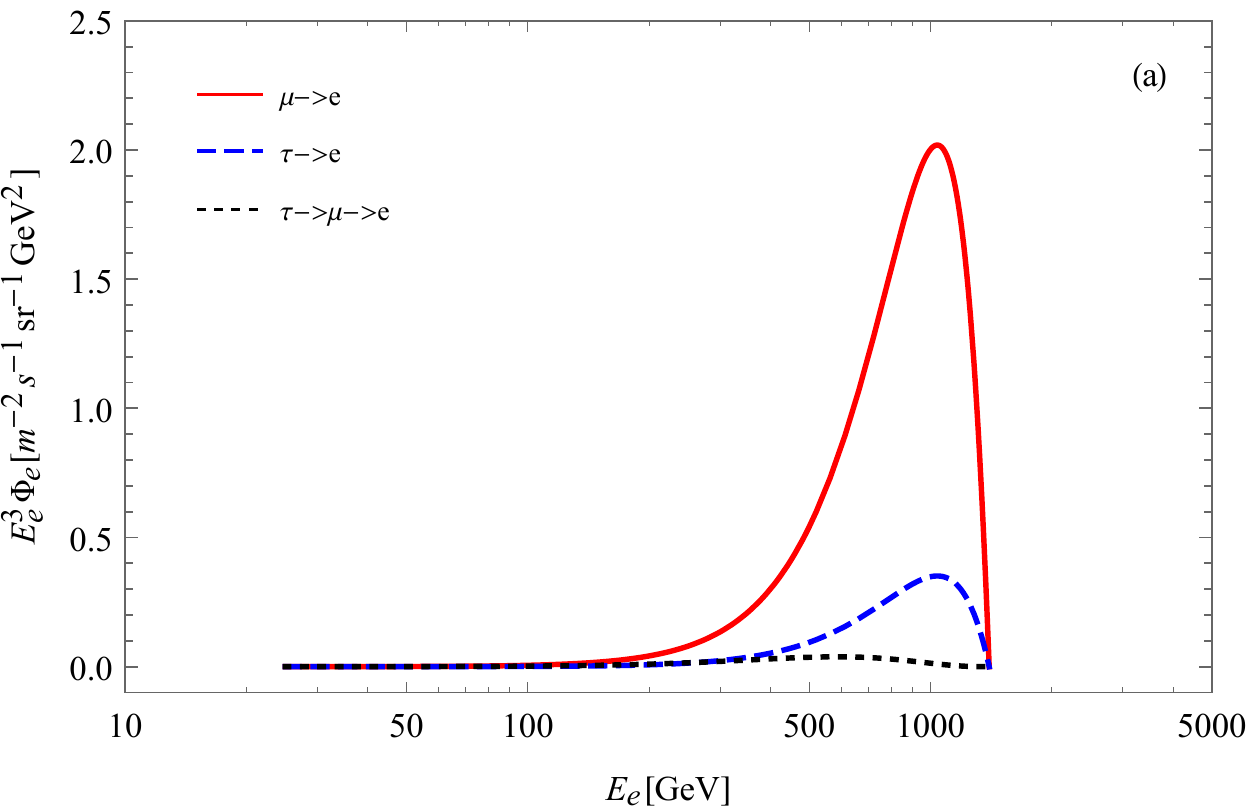}
\includegraphics[height=6.5cm,width=8.1cm]{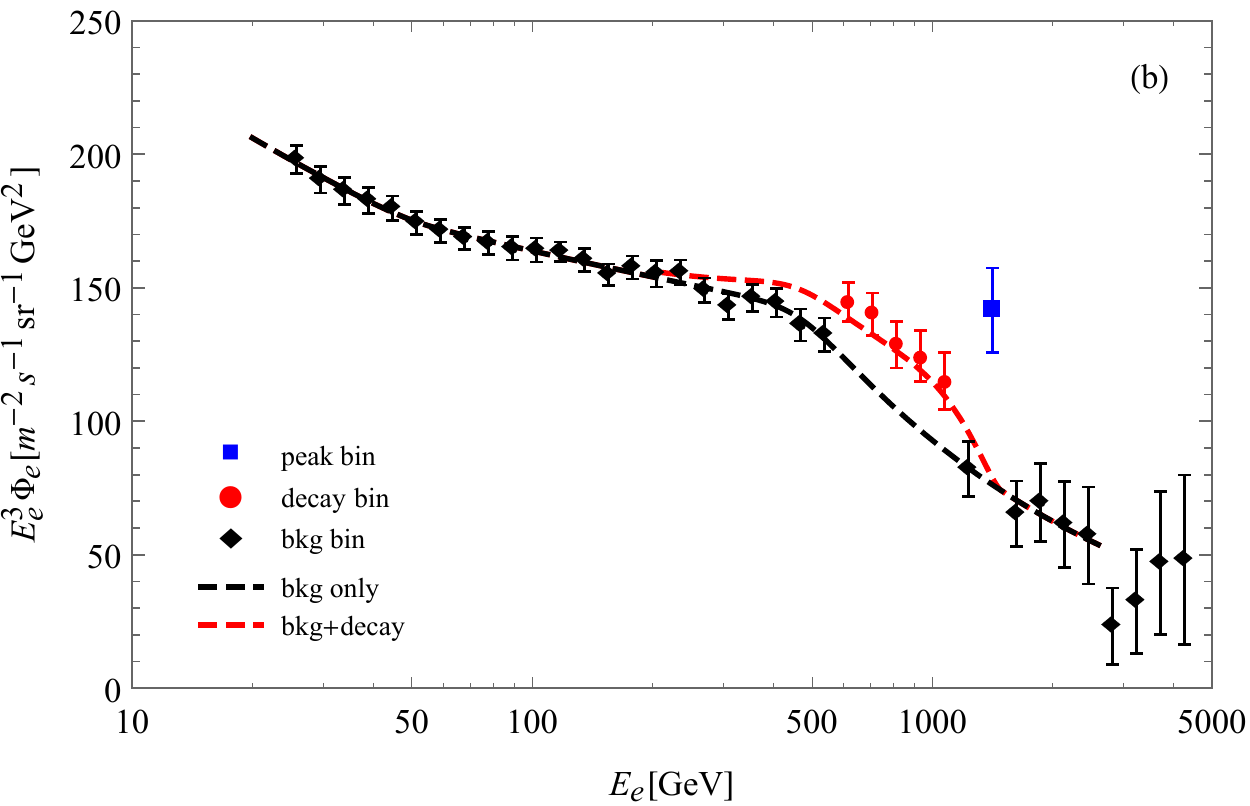}
\vspace*{-2mm}
\caption{Plot-(a):
Electron energy spectrum from the decays of 1.5\,TeV muon and tau.
The red solid curve depicts the electron energy distribution from muon decay
$\,\mu\!\to e\,\bar{\nu}_e^{}\nu_\mu^{}$,\,
the blue dashed curve shows the electron energy distribution from tau decay
$\,\tau\!\to e\,\bar{\nu}_e^{}\nu_\tau^{}$,\,
and the black dotted curve presents the electron energy distribution from
the tau decay-chain
$\,\tau\!\to\! \mu\,\bar{\nu}_\mu^{}\nu_\tau^{} \!\to\!
 e\,\bar{\nu}_e^{}\nu_\mu^{}\bar{\nu}_\mu^{}\nu_\tau^{}$.\,
Plot-(b): Fitting the background spectrum
(black bins) together with the non-peak excess (red bins) by including
decays of the 1.5\,TeV muon composition, shown by the red dashed curve.
}
\label{fig:2}
\end{figure*}

The $e^\pm$ flux from muon decay is given by
$\,\Phi_{\mu\to e}^{} = N_\mu \fr{1}{\Gamma}\fr{d \Gamma}{d E_e^{}}$,\,
as a product of the muon event number $N_\mu$ and the
normalized $e^\pm$ spectrum, which we compute as follows,
%
\beqa
\label{eq:muDecay}
\hspace*{-2.5mm}
  \frac 1 \Gamma \frac {d \Gamma}{d E_e}
\simeq
  \frac 4 {\,E_\mu} \!
\left(
  \frac 5 {12}
- \frac {3 E^2_e}{\,4 E^2_\mu\,}
+
  \frac {E^3_e}{\,3 E^3_\mu\,}
\right) ,
\eeqa
%
for $\,E_e^{}, E_\mu^{}\gg m_e^{},m_\mu^{}$.\,
The $E^3_e$ weighted $e^\pm$ energy spectrum
is shown as the red solid curve in \gfig{fig:2}(a).
In this plot, we use the exact decay formula for computation and
find no visible difference from using the above Eq.\eqref{eq:muDecay}.
The major contribution to $E^3_e \Phi_e$ appears around $(0.6-1.1)$TeV
which well match the new excess region we identified in Sec.\,\ref{sec:2}.
Hence, the $1.5\,$TeV muon decay products of $e^\pm$ are truly promising
to explain this new excess.

\vspace*{1mm}

The $\tau^\pm$ lepton can decay into $e^\pm$ via two channels,
the single 3-body-decay
$\,\tau\to e \,\bar{\nu}^{}_e\nu_\tau^{}\,$
and the chain decay
$\,\tau\to \mu \bar{\nu}^{}_\mu\nu_\tau^{}
       \to (e\,\bar{\nu}_e^{}\nu_\mu^{})\bar{\nu}^{}_\mu\nu_\tau^{} \,$.
Since each decay process is mediated by the $W^\pm$ bosons
with exactly the same gauge coupling and the lepton masses are negligible
as compared to the initial lepton energy $1.5$\,TeV, the decay
$\,\tau\to e \,\bar{\nu}^{}_e\nu_\tau^{}\,$ has almost the same $e^\pm$
spectrum as the muon decay. The major difference comes from their decay
branching fractions\,\cite{PDG},
Br$[\mu\!\to\! e\,\bar{\nu}_e^{}\nu_\mu^{}]\simeq 100\%$
and
Br$[\tau\!\to\! e \,\bar{\nu}^{}_e\nu_\tau^{}]\simeq 17.83\%\simeq 1/5.6$\,.\,
In \gfig{fig:2}(a), we present the $e^\pm$ spectrum from
the 3-body-decays of muon by the red solid curve
and that of tau by the blue dashed curve.
Also, the decay $\,\tau\!\to\! \mu\,\bar{\nu}^{}_\mu\nu_\tau^{}$\,
has a branching fraction $17.4\%$.
For the chain decay, the $e^\pm$ comes
from the secondary decay product $\mu^\pm$ and thus has much lower energy.
Consequently, its contribution to $E^3_e \Phi_e^{}$ is highly suppressed,
as shown by the black dotted curve in \gfig{fig:2}(a). Hence, the major contribution
of tau decay comes from $\,\tau\!\to\! e \,\bar{\nu}^{}_e\nu_\tau^{}\,$,\,
which has a similar electron energy distribution to that of the muon decay,
but with a suppression factor $\sim\!\fr{1}{6}$.\, This means that
the decay contributions of muon and tau to the DAMPE CRE spectrum are highly
degenerate. For the given $(\mu,\,\tau)$ event numbers $(N_\mu,\, N_\tau)$,\,
their total contribution is equivalent to
$\,(N_\mu \!+\! \fr{1}{6}N_\tau)$ number of muon decay events.
So the decay contribution of a possible tau component to the CRE spectrum is minor.
The simplest realization is that all decay contributions arise from
the muon events.

\vspace*{1mm}

For comparison, in \gfig{fig:2}(a) we set the number of muon and tau events
be the same as the peak electron events. Although the peak
data in \gfig{fig:1} is much higher than the background curve (with a
difference around $\,70$\, units for $E^3_e \Phi_e^{}$),\,
the decay spectrum is much lower.
The difference comes from the fact that the peak bin is around
$1.4\,$TeV, but the decay peak is around $1\,$TeV.
The weight factor $E^3_e$ can lead to an enhancement factor
$\,1.4^3 \!\sim\! 2.7$\, in the plot.
Furthermore, the electron excess around $1.4\,\mbox{TeV}$ is distributed
within a single bin, while the decay spectrum spreads a much wider region.
Hence, to explain this new excess requires much more
$\mu^\pm$ events than the $e^\pm$ events in the peak bin.
Adding the muon decay spectrum to the background of fit-A,
$\,\Phi_e^{} \equiv \Phi_{\rm bkg} \!+\! \Phi_{\mu\to e}^{}$,\,
we fit all data points and obtain the thermally averaged annihilation
cross section for muon production
$\,\left\langle \sigma v \right\rangle_\mu
 = 1.47\!\times\!10^{-25}\text{cm}^3/\text{s}$;\,
while for the annihilation into electrons, we have
$\,\left\langle \sigma v \right\rangle_e
  = 1.72\!\times\!10^{-26}\text{cm}^3/\text{s}$\,.\,
The ratio of cross sections between the decayed muons and the peak electrons is
$\,y \equiv \left\langle \sigma v \right\rangle_\mu
\!/\!\left\langle \sigma v \right\rangle_e \simeq 8.6$\,.\,
We present our new fit of including the muon decay contribution as
the red dashed curve in \gfig{fig:2}(b).
Impressively, it demonstrates that including the muon decay events
can fully explain this non-peak-like new excess
in the energy region $(0.6\!-\!1.1)$TeV.
If we set the second break parameter 
be its 90\% upper limit $\,E_{\text{br2}}^{}=738$\,GeV,
we find that this ratio decreases to $\,y = 2.6$\,.\,
This is because a larger $E_{\text{br2}}^{}$ value tends to reduce
the area between the background curve and the 5 red bins.
When taking 90\% lower limit
$E_{\text{br2}}^{}=379$\,GeV, we obtain $\,y = 10.8$\,.\,
Thus, we have $\,y=2.6-10.8\,$ at 90\%\,C.L., and we further derive
$\,y=8.6^{+1.4}_{-2.5}\,$ with the $\pm 1\sigma$ errors.

\section{Origin of the Flavor Structure for the CRE Excesses at DAMPE}
\label{sec:4}

When cosmic-ray electrons/positrons (CRE) travel across the interstellar space,
they would experience diffusion and energy loss.
This process can be described by the following diffusion equation,
\beqa
  \frac{\,\partial \Phi_e^{}}{\partial t}
- \frac{\,\partial [b(E) \Phi_e]\,}{\partial E}
- D(E)\! \bigtriangledown^2\!\Phi_e^{}
\,=\,  Q \,,
\label{eq:diffusion}
\eeqa
where $\,\Phi_e^{} (E_e, t, {\bf x})$\, is the number density function of the $e^\pm$
energy and the spacetime coordinates.
The energy loss, $\,b(E) \equiv - dE/dt$\,,\,
can be parametrized as $\,b(E) = b_0^{}(E/\text{GeV})^2$\,
with
$\,b_0^{} = 10^{-16} \text{GeV}/\text{s}$.\,
The diffusion coefficient is
$\,D(E) = D_0 (E/\mbox{GeV})^\delta$,\, where
$\,D_0^{} = 11 \mbox{pc}^2/\mbox{kyr}$\,
and $\,\delta = 0.7$\,.\,
The right-hand-side of the above Eq.\eqref{eq:diffusion}
is the $e^\pm$ source function,
$\,Q({\bf x}, E_e) \propto \rho_\chi^2({\bf x})
\left\langle \sigma v \right\rangle {dN}/{dE_e}$\,,\,
where $\,\rho_\chi^{}({\bf x})\,$ is the DM density distribution and
$\,dN/dE_e\,$ is the $e^\pm$ energy spectrum
from the DM annihilation. Our calculation takes
the spherically symmetric NFW density profile\,\cite{NFW}
$\,\rho_\chi^{}(r) \equiv
   \rho_s^{} (r/r_s^{})^{-\gamma}(1 \!+\! r/r_s^{})^{\gamma-3}\,$
for a nearby DM subhalo.

\begin{figure*}[t]
\centering
\includegraphics[height=8cm,width=11cm]{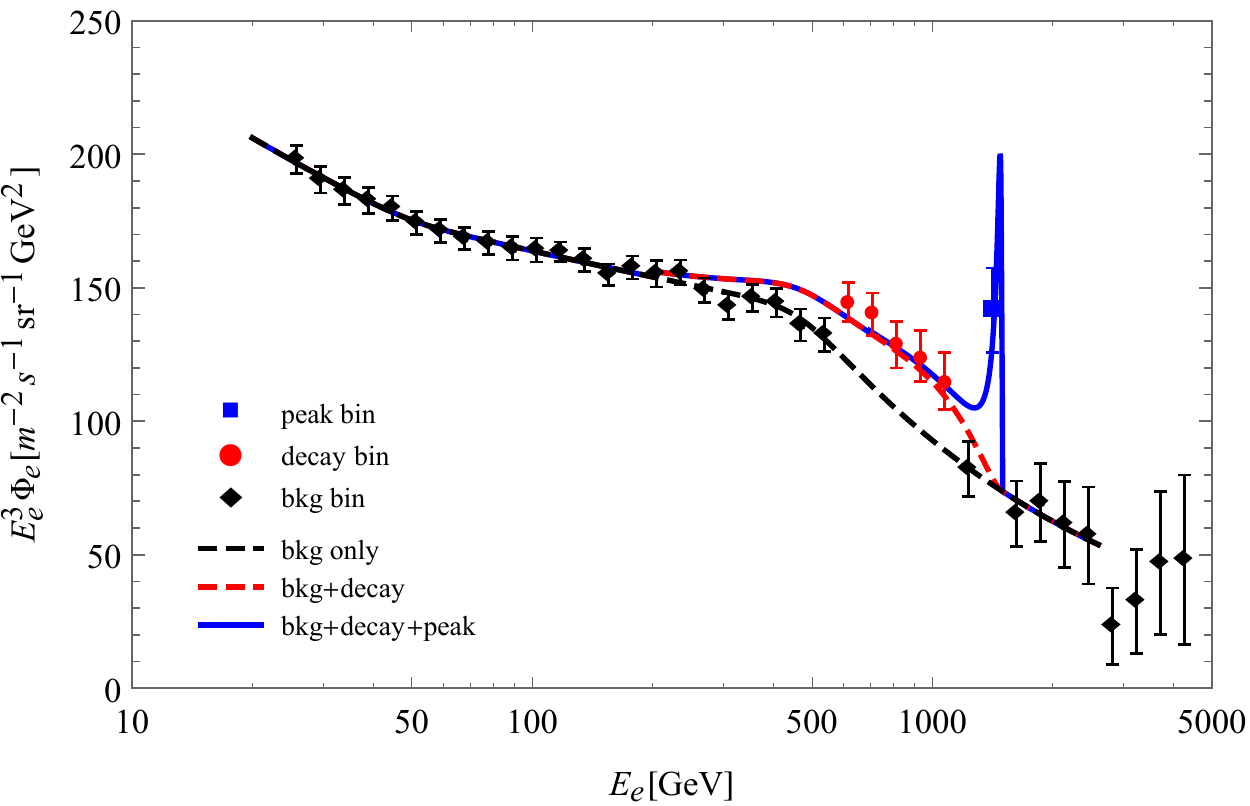}
\vspace*{-2mm}
\caption{CRE spectrum for DAMPE: including both the decay contributions of
1.4TeV $\mu^\mp$ composition (blue curve) together with the 1.4TeV $e^\mp$ peak-like
contribution (green curve), in addition to the pure backgrounds (fitted by the
black curve).}
\label{fig:fit-all}
\label{fig:3}
\end{figure*}

\vspace*{1mm}

The diffusion function \geqn{eq:diffusion} can be solved with Green function
\cite{diffusion},
\begin{equation}
  G({\bf x}, E; {\bf x}_s^{}, E_s^{}) \,=\,
  \frac{\,\exp\!\left[ - |{\bf x} - {\bf x}_s^{}|^2/\lambda^2 \right]\,}
				{b(E) (\pi \lambda^2)^{3/2}} \,,
\end{equation}
where $E_s^{}$ is the electron/positron energy at source and $E$ the counterpart
after diffusion. The propagation scale $\,\lambda\,$ is given by
$\,\lambda^2 = 4 \int^{E_s}_E\!\! d E' {D(E')}/{b(E')}$\,.\,
Then, the solution of Eq.\geqn{eq:diffusion} can be expressed as
\beqa
  \Phi_e^{}(E_e^{}) \,=
  \int\!\! d^3 x_s^{}\!\! \int \!\! d E_s^{}\,
  G({\bf x}, E_e^{}; {\bf x}_s^{}, E_s^{})
  Q({\bf x}_s^{}, E_s^{}) \,.
\eeqa

To explain the narrow peak around $(1.3\!-\!1.5)$TeV,
the energy loss cannot be large.
This can be achieved if the DM subhalo is close to the earth. For illustration,
we choose $\,\gamma = 0.5$,\, the subhalo radius $\,r_s^{}\! = 0.1$\,kpc,\,
and the subhalo distance $\,d_s^{} = 0.2\,\text{kpc}$.
Consequently, the mono-energetic $e^\pm$ peak
at 1.5\,TeV slightly spreads to lower energy and lies mainly
within the peak bin $(1.3\!-\!1.5)$TeV,
which is shown by the blue solid curve in \gfig{fig:3}.
As we mentioned earlier in Sec.\,\ref{sec:3},
the 1.5\,TeV $\mu^\pm$ events have a lifetime about 0.031s,\,
so they will decay into $e^\pm$ shortly after their production at the source.
We have further considered the diffusion effect for the muon decay contribution.
This is included in the red dashed curve
in \gfig{fig:2}(b) and \gfig{fig:3}.
We note that in \gfig{fig:2}(a) the 1.5\,TeV muon decay distribution of
$\,E^3_e \Phi_{\mu \rightarrow e}$
is shifted towards lower energy and exhibits a peak around 1.1\,TeV.

\vspace*{1mm}

In \gfig{fig:3}, based upon our background fit-A (black dashed curve),
we perform a combined fit to the new excess over $(0.6-\!1.1)$TeV and
the peak excess at $(1.3\!-\!1.5)$TeV, where the red curve shows the
muon decay contribution alone and the blue curve depicts the summed
contribution from both the 1.5\,TeV $\mu^\pm$ and 1.5\,TeV $e^\pm$ events.
In this combined fit, we obtain $\,\chi^2 =16.5$\, with $\text{d.o.f.}=27$,\,
and thus $\,\chi^2\!/\text{d.o.f.}=16.5/27\simeq 0.609$.\,
As expected, this has a lower $\chi^2$ and better fitting quality
than the naive fit of all bins in the same energy region $(0.025-2.6)$TeV
which gives $\,\chi^2\!/\text{d.o.f.}=22.8/29\!=\!0.785$
(Table\,\ref{tab:1}) and increases $\,\chi^2\,$ by $\,\Delta\chi^2\!=6.3\,$.\,
Our combined fit is also better than another naive fit over the range
$(0.055\!-\!2.6)$TeV, which has
$\,\chi^2\!/\text{d.o.f.}\!=22.5/24=0.939$\,
(Table\,\ref{tab:1}).

\vspace*{1mm}

From the analysis in \gfig{fig:2} and \gfig{fig:3}, we find
that the original lepton final state produced at a nearby source
should have the flavor composition ratio,
\beqa
N_e : \(N_\mu \!+\!\fr{1}{6}N_\tau\) \,=\, 1:y\,,\,
\eeqa
with $\,y=8.6^{+1.4}_{-2.5}\,$,\,
or, $\,y=2.6-10.8\,$ at 90\%\,C.L.
Note that the $\tau$ component could only play a minor role here
due to the suppression factor $\sim\!\fr{1}{6}$\, by its small
decay branching fraction of
$\,\tau\!\to\!e\,\bar{\nu}_e^{}\nu_\tau^{}\,$.\,
The simplest realization of this flavor composition condition is
\,$N_e : N_\mu : N_\tau = 1 : y:0$\,.\,

\vspace*{1mm}

The above flavor composition condition will place important
constraint on the lepton-related DM model buildings.
For instance, for the typical lepton portal
DM models\,\cite{DAMPE-imply}\cite{LPDM},
the DM can be either a fermion or scalar.
In the first case, a neutral singlet Dirac fermion $\chi$ can serve as DM
and couples to a scalar mediator $S$ and the right-handed charged lepton
$\ell_{Rj}^{}$\,,
\beqa
\label{eq:DM-fermion}
{\cal L}_{\chi}^{} \,\supset\,
\lambda_j^{}S_j^{}\,\overline{\chi_L^{}}\ell_{Rj}^{} + \text{h.c.},
\hspace*{6mm}
\eeqa
where $\,\ell_j^{}=e,\mu,\tau$.\,
In the second case, the DM particle $X$ is a neutral
complex singlet scalar and the mediator $\psi$ is a Dirac fermion
with the same electric charge and lepton number as the charged leptons.
The Lagrangian contains the relevant interaction vertex
\beqa
\label{eq:DM-scalar}
{\cal L}_{X}^{} \,\supset\,
\la_j^{}X\overline{\psi_{Lj}^{}}\ell_{Rj}^{}+\text{h.c.}
\hspace*{6mm}
\eeqa

\vspace*{1mm}

Thus, in the above models, the DM annihilation $\chi\bar{\chi}\to \ell_j^{}\ell_j^{}$
or $XX\to \ell_j^{}\ell_j^{}$
goes through the $t$-channel exchange of $S_j^{}$ or $\psi_{Lj}^{}$.\,
The annihilation cross section is proportional to
$\,\lambda_j^4\,$.\,
Thus, our simplest realization of the flavor composition condition gives
$\,N_e \!:\! N_\mu \!:\! N_\tau
 = \lambda_e^4\!:\! \lambda_\mu^4 \!:\! \lambda_\tau^4
 = 1\!:\! y\!:\!0$\,,\, with $\,y=2.6-10.8\,$ at 90\%\,C.L.
This means that the DM coupling to $\tau$ leptons
is forbidden. Thus, we deduce a simple coupling relation,
$\,\lambda_e^{}\!:\! \lambda_\mu^{} \!:\! \lambda_\tau^{}
  = 1\!:\! y^{\frac{1}{4}}\!:\!0$\,,\,
  with $\,y^{\frac{1}{4}} \simeq 1.3\!-\!1.8\,$.\,
More generally, we have the following coupling condition,
\beqa
\label{eq:mutau}
\la_e^{} : \(\la_\mu^4 + \fr{1}{6}\la_\tau^4\)^{\!\fr{1}{4}}
=\, 1:y^{\frac{1}{4}}\,,\,
\eeqa
where $\,y^{\frac{1}{4}}=1.3-1.8\,$ (90\%\,C.L.)
is a fairly narrow range.
If a $\mu-\tau$ flavor symmetry requires $\la_\mu^{}\!=\la_\tau^{}$,\,
then we find that Eq.\eqref{eq:mutau} further leads to a simple coupling constraint,
$\,\lambda_e^{}\!:\! \lambda_\mu^{} \!:\! \lambda_\tau^{}
 = 1\!:\! \tilde{y}^{\frac{1}{4}}\!:\!\tilde{y}^{\frac{1}{4}}$,\,
with
$\,\tilde{y}^{\frac{1}{4}}
 \!=\!\(\frac{6}{7}y\)^{\frac{1}{4}}\!\simeq\! 1.2\!-\!1.7$.
Further applications to the DM model buildings
are encouraging and will be pursued elsewhere.

\section{Conclusions}
\label{sec:5}

The new announcement of detecting the TeV cosmic-ray electrons/positrons (CRE)
by the DAMPE collaboration\,\cite{DAMPE2017} has brought up further excitements
for probing the nearby galactic sources and possible dark matter (DM)
annihilations.

\vspace*{1mm}

In Section\,\ref{sec:2}, we inspected the DAMPE CRE
energy spectrum\,\cite{DAMPE2017}
and uncovered a new hidden excess of non-peak-like structure over the
region $(0.6-\!1.1)$TeV, which is shown by the 5 red bins in Fig.\,\ref{fig:1}.
This non-peak-like new excess is significant as compared to
the well-noticed peak excess around 1.4\,TeV.
We performed various fits of the background bins in Fig.\,\ref{fig:1}
by using either the double broken power-law formula \eqref{eq:DBPL}
or the single broken power-law formula \eqref{eq:SBPL} (adopted by
the DAMPE group\,\cite{DAMPE2017}).
We treat the new excess (5 red bins) and the peak excess (blue bin)
as signals and fit the rest of data points (black bins) for the
background curve. We found that these new fits have much improved
qualify of $\chi^2$ fitting as compared to the naive fit of treating all bins
as the backgrounds. A summary of our background fitting analysis is presented
in Table\,\ref{tab:1}. This demonstrates that our background fits
are well motivated and instructive.

\vspace*{1mm}

In Section\,\ref{sec:3}, given the encouraging observation in Fig.\,\ref{fig:1},
we proposed a new mechanism to explain this non-peak-like new excess.
We conjectured that this new excess originates from
decays of the final state $\mu^\pm$ (and $\tau^\pm$)
produced by the 1.5\,TeV DM annihilations (or other mechanism),
in addition to the final state $e^\pm$ which explain the peak excess.
We analyzed the $e^\pm$ energy distributions from both muon
and tau decays. We found that muon decays always give the dominant contribution,
as shown by the red solid curve in Fig.\,\ref{fig:2}(a).
Then, we included the muon decay contribution into the fit of the
DAMPE CRE spectrum, and demonstrated that this can fully explain the new
excess over the $(0.6-1.1)$TeV energy region,
as shown in Fig.\,\ref{fig:2}(b).

\vspace*{1mm}

In Section\,\ref{sec:4}, we further performed a combined fit to the DAMPE CRE spectrum
by including both the decay contribution from the 1.5\,TeV $\mu^\pm$ events
and the peak contribution from the 1.5\,TeV $e^\pm$ events.
For illustration, we included the diffusion effects on the
CRE spectrum after the produced $e^\pm$ events propagate from
a nearby galactic source at
a distance of about 0.2\,kpc to the DAMPE detector.
We presented our fit to the full CRE spectrum including
the decay and peak contributions together in Fig.\,\ref{fig:3}.
This impressively explains both the non-peak-like new excess
over the $(0.6\!-\!1.1)$TeV region and the peak excess at
$(1.3\!-\!1.5)$TeV.

\vspace*{1mm}

Our analysis demonstrated that the {\it flavor structure} of the original
lepton final-state produced by the DM annihilations (or other mechanism)
in a nearby clump or subhalo should have a flavor composition ratio
$N_e \!:\! (N_\mu \!+\! \frac{1}{6} N_\tau) \!= 1\!:\!y$\,
with $y= 8.6^{+1.4}_{-2.5}$ or
$\,y\simeq 2.6\!-\!10.8$ at 90\%\,C.L.
For lepton portal DM models,
this imposes a nontrivial bound on the lepton-DM-mediator couplings
$\,\lambda_e^{} \!:\! (\lambda_\mu^4 \!+\!\frac{1}{6}\lambda_\tau^4)^{\frac{1}{4}}
 =1\!:\!y^{\frac{1}{4}}$,\,
with a narrow range $\,y^{\frac{1}{4}}\simeq 1.3\!-\!1.8\,$ (90\%\,C.L.).
Such constraints are important for the DM model buildings.

\vspace*{1mm}

We further note that this non-peak-like new excess over $(0.6\!-\!1.1)$TeV
is also consistent with the recent data of another satellite experiment
Fermi-LAT\,\cite{Fermi}
(which are shown in Fig.\,2 of Ref.\,\cite{DAMPE2017} for comparison).
Although confirming the new excess and/or the tentative 1.4\,TeV peak
structure would require more data-taking from DAMPE and
other CRE experiments, these encouraging clues are important and
deserve further investigations.
It is worth to stress that
the continued DAMPE runs up to 6 years will reduce the statistical errors
by half and help to pin down the new excess over $(0.6\!-\!1.1)$TeV
in addition to the 1.4\,TeV peak structure,
as shown in Table\,\ref{tab:1}.

\vspace*{4mm}
\noindent
{\bf Acknowledgments}:
\\[1mm]
We thank Xiaojun Bi, Mingshui Chen,  Hong-Bo Hu, Qiang Yuan, Yue Zhao and Yu-Feng Zhou
for useful discussions.
This research was supported in part by the national NSFC
(under grants 11675086 and 11275101),
by the Shanghai Key Laboratory for Particle Physics and Cosmology
(under grant 11DZ2260700),  by the Office of Science and Technology,
Shanghai Municipal Government (under grant 16DZ2260200),
by the MOE Key Laboratory for Particle Physics, Astrophysics and Cosmology,
and by the CAS Center for Excellence in Particle Physics (CCEPP).
This work was also supported by World Premier International Research
Center Initiative (WPI Initiative), MEXT, Japan.

\end{document}